\documentclass[iop]{emulateapj}

\usepackage{apjfonts}
\usepackage{pbox}
\usepackage{bm}
\usepackage{CJK}
\usepackage{color}

\shorttitle{Measuring O abundances from spectra without O lines}
\shortauthors{Ting et al.}

\begin{document}

\begin{CJK*}{UTF8}{gbsn}
\title{Measuring oxygen abundances from stellar spectra without oxygen lines}
\author{Yuan-Sen Ting (丁源森)\altaffilmark{1,2,3,4}, Charlie Conroy\altaffilmark{5}, Hans-Walter Rix\altaffilmark{6}, Martin Asplund\altaffilmark{3}}
\altaffiltext{1}{Institute for Advanced Study, Princeton, NJ 08540, USA}
\altaffiltext{2}{Department of Astrophysical Sciences, Princeton University, Princeton, NJ 08544, USA}
\altaffiltext{3}{Research School of Astronomy and Astrophysics, Australian National University, Cotter Road, ACT 2611, Canberra, Australia}
\altaffiltext{4}{Observatories of the Carnegie Institution of Washington, 813 Santa Barbara Street, Pasadena, CA 91101, USA}
\altaffiltext{5}{Harvard--Smithsonian Center for Astrophysics, 60 Garden Street, Cambridge, MA 02138, USA}
\altaffiltext{6}{Max Planck Institute for Astronomy, K\"onigstuhl 17, D-69117 Heidelberg, Germany}
\slugcomment{Submitted to ApJ}

%
%
%
%
%
%
\begin{abstract} 
Oxygen is the most abundant ``metal'' element in stars and in the cosmos. But determining oxygen abundances in stars has proven challenging, because of the shortage of detectable atomic oxygen lines in their optical spectra as well as observational and theoretical complications with these lines (e.g., blends, 3D, non-LTE). Nonetheless, \citet{tin17b} were recently able to demonstrate that oxygen abundances can be determined from low-resolution ($R \simeq 2000$) optical spectra. Here we investigate the physical processes that enable such a measurement for cool stars, such as K-giants. We show that the strongest spectral diagnostics of oxygen come from the CNO atomic-molecular network, but are manifested in spectral features that do not involve oxygen.  In the outer atmosphere layers most of the carbon is locked up in CO, and changes to the oxygen abundance directly affect the abundances of all other carbon-bearing molecules, thereby changing the strength of CH, CN, and C$_2$ features across the optical spectrum. In deeper atmosphere layers most of the carbon is in atomic form, and any change in the oxygen abundance has little effect on the other carbon-bearing molecules. The key physical effect enabling such oxygen abundance measurements is that spectral features in the optical arise from both the CO-dominant and the atomic carbon-dominant regions, providing non-degenerate constraints on both C and O. Beyond the case at hand, the results show that physically sound abundances measurements need not be limited to those elements that have observable lines themselves. 
\end{abstract}

\keywords{methods: data analysis --- stars: abundances }

%
%
%
%
%
%

\section{Introduction}
\label{sec:introduction}

Oxygen is the third most abundant element in the cosmos, and the most abundant ``metal" and plays a key role in both stellar and galactic evolution. Almost all oxygen is ejected by core-collapse supernovae \citep{tim95,kob06}, whose progenitors have a short lifetime. Oxygen abundances are therefore sensitive to short timescale behavior in galactic chemical evolution. Measuring precise and accurate abundances of oxygen is critical for understanding the complex and puzzling stellar populations within globular clusters \citep[e.g.,][]{car09,mar11}. Due to the importance of the CNO cycle and as a source of opacity \citep[e.g.,][]{dot07,van12}, oxygen plays a crucial role in determining stellar structure, stellar evolution and therefore the ages of stars \citep[e.g.,][]{kra03,bon13}. Moreover, determining oxygen abundances for exoplanet host stars is important since the C/O ratio can affect the atmospheres and habitability of exoplanets \citep[e.g.,][]{mad12}.

Finally, CO is commonly adopted as a tracer of molecular gas since molecular hydrogen is hard to measure \citep[see][for a review]{bol13}. It is also the most readily measured element abundance in the ionized ISM, from the Milky Way to the highest redshifts. Measuring oxygen in stars is hence a direct way of comparing the enrichment of stars and of the ISM.

\begin{figure*}
\centering
\includegraphics[width=1.0\textwidth]{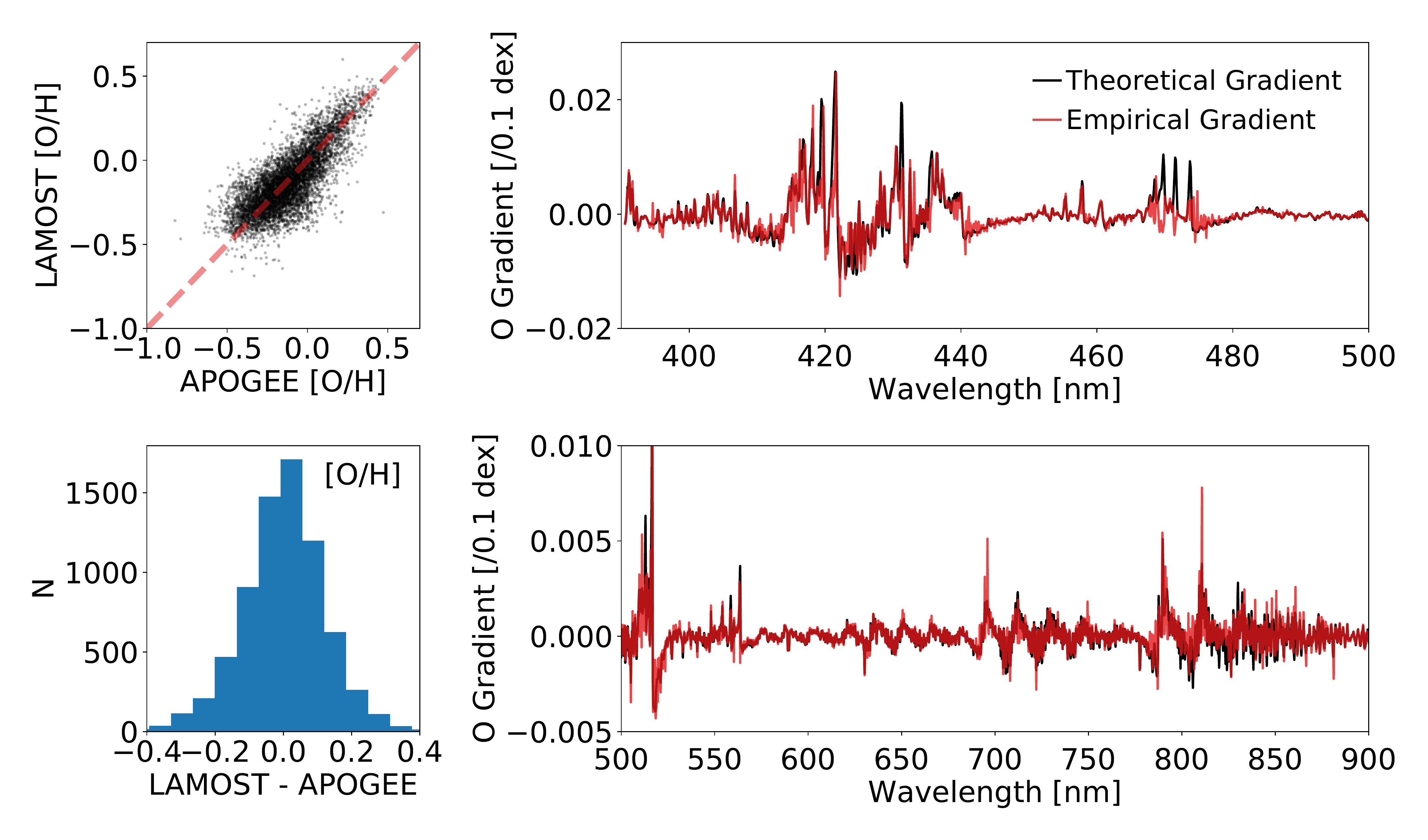}
\caption{Measuring oxygen abundances from low-resolution LAMOST spectra at $R \simeq 2000$. The top left panel shows the comparison between oxygen abundances derived from LAMOST and from APOGEE spectra. Their difference characterized in the bottom left panel -- the oxygen measurements have a scatter of $\sigma \approx 0.1\,$dex. The right panels compare the oxygen response functions from the \citet{tin17b} data-driven models to those from theoretical models. The agreement indicates that our method infers oxygen abundances from features predicted by the theoretical models.}
\label{fig1}
\end{figure*}

\begin{figure*}
\hspace{-1cm}
\includegraphics[width=1.05\textwidth]{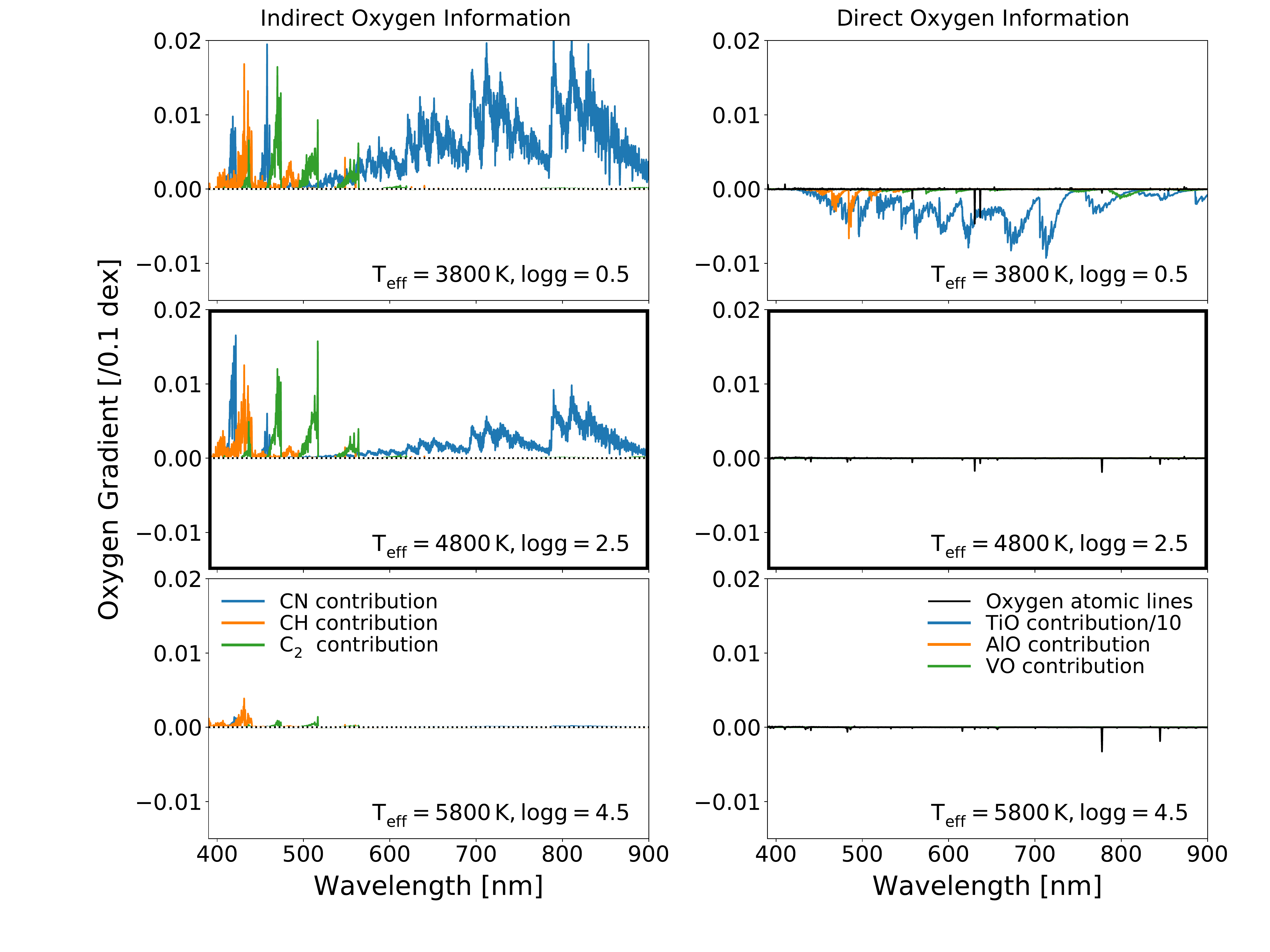}
\caption{Theoretical oxygen response function of optical spectra at $R=2000$, showing how much a spectrum varies as oxygen abundance increases by 0.1 dex, with all other labels held fixed. The top, middle and bottom panels show the response functions for M-giants ($T_{\rm eff} = 3800\,{\rm K}$, $\log g = 0.5$), K-giants ($T_{\rm eff} = 4800\,{\rm K}$, $\log g = 2.5$), and G-dwarfs ($T_{\rm eff} = 5800\,{\rm K}$, $\log g = 4.5$), respectively, all with solar metallicity. The middle panels also correspond to the $T_{\rm eff}$/$\log g$ combination explored in other figures. The left panels show indirect oxygen information that is not from oxygen species, while the right panels show oxygen information that is directly from oxygen species. For K-giants, atomic oxygen features (black lines) in the right panels are too weak to be reliably measured at LAMOST resolution, and there is no other direct oxygen molecular feature. However, changing oxygen abundance affects the CO abundance in the atmosphere which in turn indirectly modifies the abundances of CH, CN, and ${\rm C}_2$. At cooler temperatures, TiO contributes significantly to the spectra which allows a direct measurement of oxygen. Note that the TiO contribution is divided by 10 for display purposes. AlO and VO also contribute for the cooler stars but their contributions are negligible compared to TiO. At higher temperatures (bottom panels), the indirect oxygen information becomes much weaker.}
\label{fig2}
\end{figure*}

Despite its importance, it is difficult to measure oxygen abundances of stars \citep[e.g.,][]{asp05}. Most ongoing large-scale spectroscopic surveys, such as GALAH \citep{mar17}, Gaia-ESO \citep{gil12}, RAVE \citep{kun17}, and LAMOST \citep{xia17b}, are designed to work in the optical, which is rich in spectral abundance diagnostics \citep[see appendix of][for a summary]{tin17a}. But there are only a few oxygen lines in the optical. Moreover, the atomic oxygen transitions are either strongly blended with other lines (even at high resolution) or suffer non-negligible non-local thermodynamic equilibrium (NLTE) effects \citep[see a review from][]{asp09}. Alternatively, one can exploit information from oxygen-bearing molecules such as OH or TiO. The former molecule displays many transitions in the infrared, which has been exploited by the APOGEE survey to measure oxygen abundances \citep{gar16}. TiO contains many transitions in the optical but is only present in very cool stars (e.g., $\lesssim 4200{\rm K})$ and is therefore not a useful oxygen indicator for the bulk of the samples in existing spectroscopic surveys. In light of these issues, deriving an independent diagnostic of the oxygen abundance in stars central to the large Galactic surveys (e.g., K-giants) would be very valuable. Ideally such oxygen measurements should be possible with optical spectra, and also at low spectral resolution;  this is the goal of this Letter.

%
%
%
%
%
%

\section{Oxygen spectral information in the optical}
\label{sec:spectrum}

Low-resolution spectra at $R \simeq 2000$ were thought to be far less informative about element abundances than high-resolution spectra because their spectral features are blended, with perhaps only $\lesssim 3$ elemental abundances measurable from such spectra. \citet{ho17b,tin17a} demonstrated that low-resolution spectra are in fact information-rich and can be employed to derive reliable elemental abundances. \citet{tin17b} examplified this in practice by measuring 14 elemental abundances, {\it including oxygen}, from $R=1800$, ${\rm S/N}_{\rm pix} \geq 30$ LAMOST spectra. We refer interested readers to the papers for details, but we will summarize the relevant information below. The LAMOST spectra cover $400\,{\rm nm}-900\,{\rm nm}$. This wavelength range does not contain any oxygen lines clearly identifiable at this resolution at the typical LAMOST S/N; therefore, we cannot derive oxygen abundances in the usual way from direct oxygen transitions. We continuum normalize all LAMOST spectra adopting the same method as in  \citet{ho17b}: we divide the spectra by the smoothed version of the spectra with a kernel size of $50\,$\AA. As discussed in \citet{tin17b}, different continuum normalization methods do not qualitatively change the results as long as we continuum normalize all spectra in the same manner. \citet{ho17b} and \citet{tin17b} also verified that the fiber-to-fiber and other instrumental variations are not dominant systematic factors.

\begin{figure*}
\includegraphics[width=1.0\textwidth]{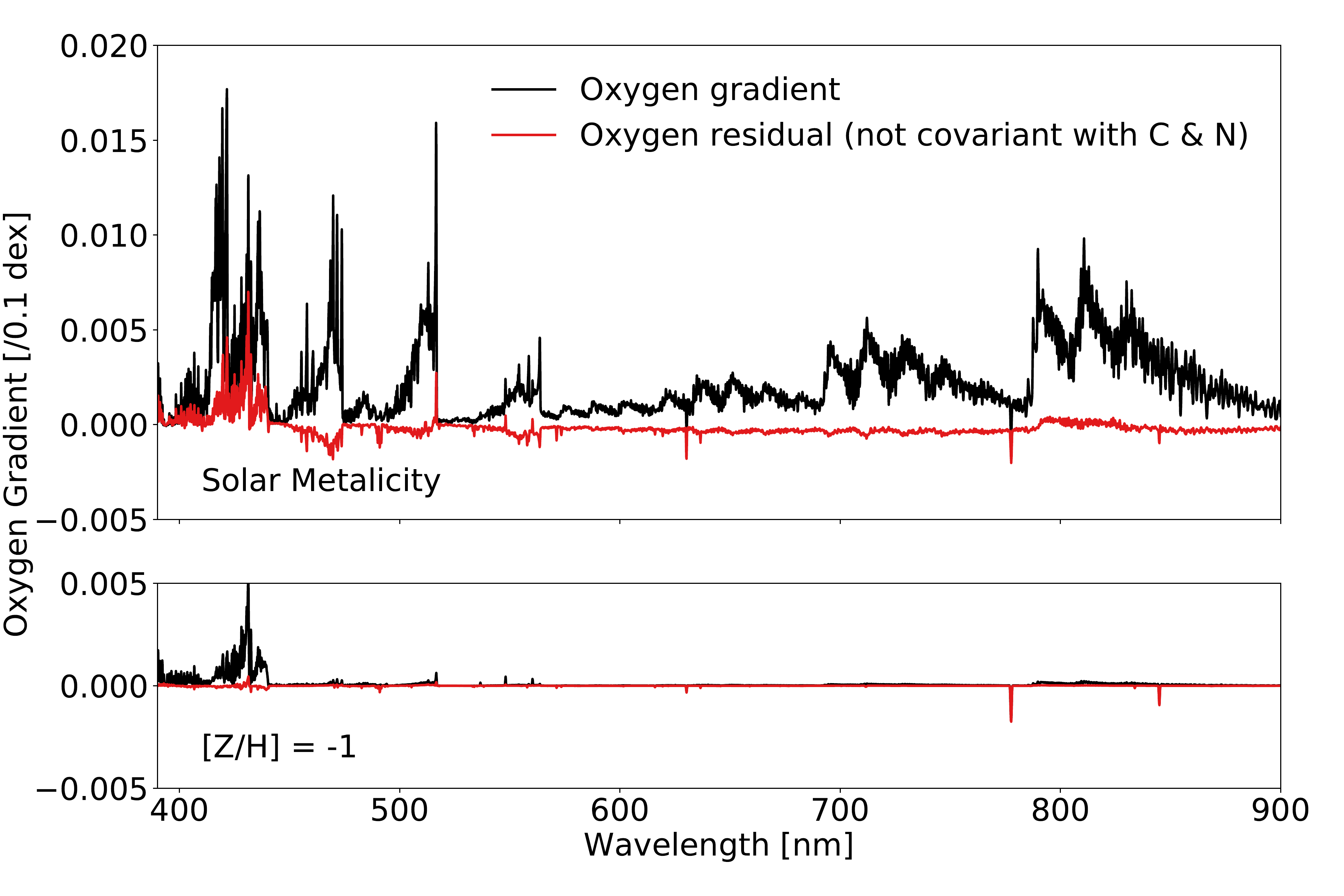}
\caption{Breaking the degeneracy of C, N, and O abundances. The black lines show the oxygen response function for a K-giant at solar metallicity (top panel) and ${\rm [Z/H]} =-1$ (bottom panel). The red lines show the remaining oxygen response function after subtracting the best combination of carbon and nitrogen response functions. At solar metallicity, the oxygen information cannot be fully mimicked by carbon and nitrogen response functions, demonstrating that the oxygen abundance can be measured {\it via} carbon-bearing molecular features.  At lower metallicity, this effect is greatly diminished, suggesting that it would be much more difficult to measure oxygen in this way at low metallicity.}
\label{fig3}
\end{figure*}

In \citet{tin17b}, we devised and applied a hybrid of data-driven \citep[e.g.,][]{nes15a,cas16,ho17b,ho17a} and theoretical models. In particular, we adopted a subset of LAMOST stars with high S/N LAMOST spectra that also have APOGEE spectra and corresponding estimates for stellar labels (elemental abundances and stellar parameters). These LAMOST high-quality counterparts are adopted as ``template'' spectra. We model the variation of the LAMOST flux at each pixel as a function of the APOGEE stellar labels using neural networks as interpolation functions. At the same time, we also enforce that the variation of the flux with each label of the data-driven models has to agree with the theoretical expectation from the synthetic models. Specifically, we require that when differentiating the data-driven models with respect to individual stellar labels, the spectral ``gradients'' (here referred to as response function) agree with the theoretical expectation, but we allow the exact strength of the response functions to be self-corrected by the data. After such a model is trained, at each new label point, we can generate a mock spectrum that can then be used to compare with the remaining observed spectra that are not used in the training. Stellar labels for those stars are inferred through $\chi^2$ minimization.

Fig~\ref{fig1} shows the oxygen abundances derived from LAMOST spectra of about 7000 APOGEE DR13-LAMOST DR3 overlapping targets. The sample consists of stars in the range $4000\,$K$<T_{\rm eff}<5000\,$K and $-0.75<{\rm [Fe/H]}<0.5$. The top left panel compares APOGEE oxygen abundance estimates against our estimates using LAMOST spectra. As shown in the bottom left panel, our LAMOST estimates agree well with the APOGEE values with a scatter of $\sigma \approx 0.1$ dex. Even with a signal-to-noise and spectral resolving power that are 10 times lower than for the APOGEE spectra, we achieve a precision that is only about 2 times worse than the APOGEE formal uncertainties \citep{hol15,sds16}. As explained in \citet{tin17b}, this agreement does not immediately imply that we have determined oxygen abundances from physically relevant spectral features, because elemental abundances can be, and often are, astrophysically correlated due to Galactic chemical evolution. By adding theoretical priors to the data-driven models \citep{tin17b}, we can ensure that we determined all abundances from the features that theoretical models predict to be spectral diagnostics, as reflected in their oxygen response function, $\partial{f_\lambda}/\partial{{\rm [O/H]}}$. To illustrate this, the right panels demonstrates the oxygen information content at the LAMOST resolution and wavelength range, comparing the oxygen response function of the data-driven model to that of the theoretical models for a typical K-giant with $T_{\rm eff} = 4800\,{\rm K}$, $\log g = 2.5$ and solar metallicity. The data-driven response function agrees very well with the theoretical expectation, demonstrating that we indeed infer oxygen abundances from the spectral features predicted to be diagnostic for oxygen by the theoretical models. Although not shown, we also checked stars that are O-rich in APOGEE, relative to a simple C-O relation, also tend to be O-rich in our LAMOST estimates, and vice versa for the O-poor stars. This further confirms that we have indeed measured oxygen abundances from oxygen spectral features, instead of exploiting astrophysical correlations between the two elemental abundances.

\begin{figure*}
\centering
\includegraphics[width=1.0\textwidth]{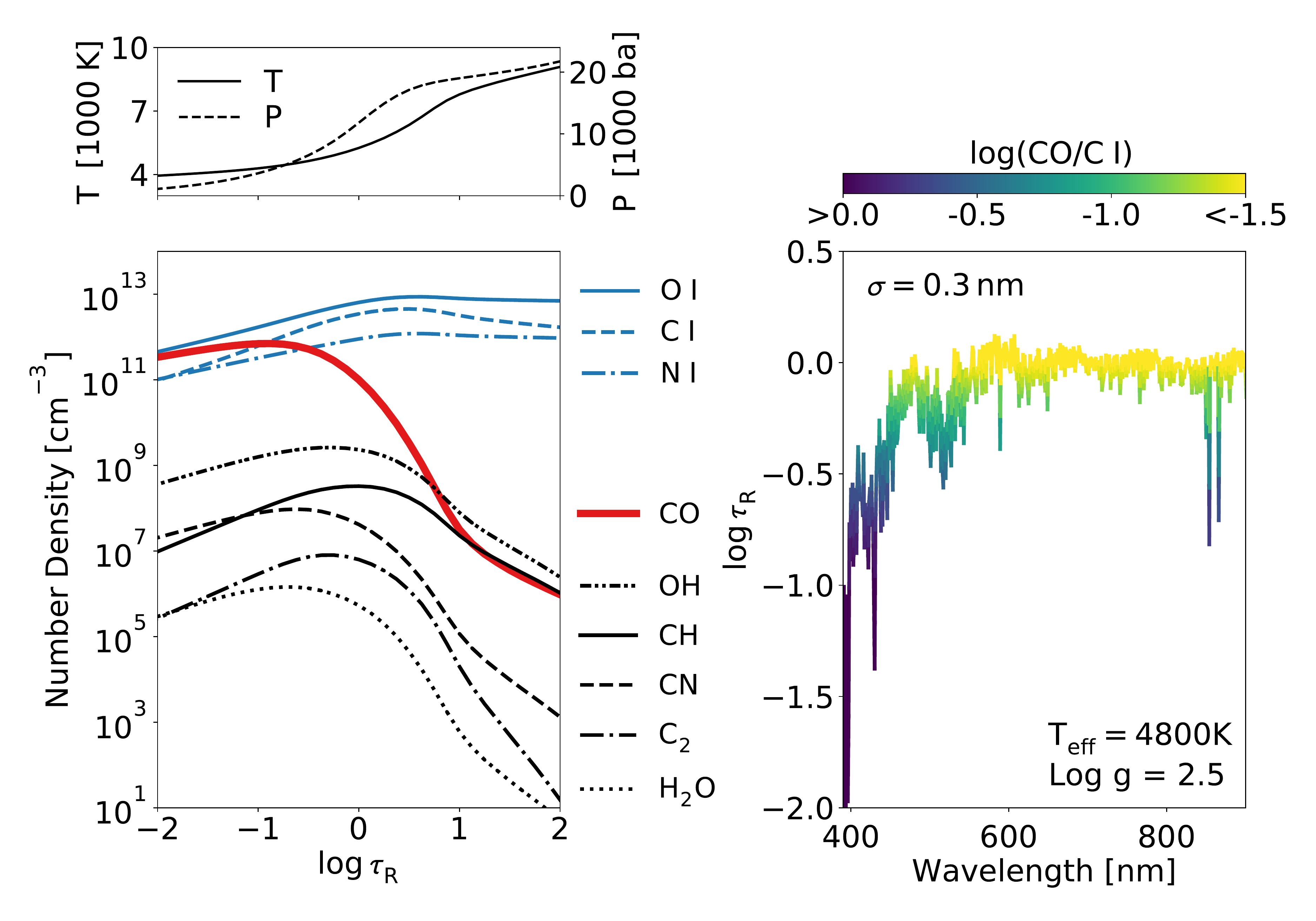}
\caption{Physical characteristics of a solar metallicity K-giant star ($T_{\rm eff}=4800$K, logg$=2.5$). {\em Left panel:} Temperature, pressure, and number density of selected CNO species as a function of Rosseland opacity. Notice the strong depth-dependence of the CO number density. {\em Right panel:} Line formation depths (where $\tau_\lambda=1$) in units of the Rosseland opacity, smoothed to 0.3$\,$nm. At bluer wavelengths most lines form in the higher atmospheric layers ($\log \,\tau_R = -2 $ to $-1$), while redder wavelengths lines form at deeper layers ($\log \,\tau_R = 0$). This figure demonstrates that optical spectra are sensitive to a range of atmospheric layers over which the dominant carbon-bearing species varies between CO and atomic carbon (as indicated by the color-coding in the right panel).}
\label{fig4}
\end{figure*}

Having shown practically that oxygen abundances can be measured without detectable oxygen transitions, we will lay out a physical explanation for why this works. Throughout this study, we adopt theoretical models generated from Kurucz models \citep[][and reference therein]{kur70, kur81, kur93, kur05, kur13, kur17} to provide a qualitative, intuitive explanation. Similar to \citet{tin17b}, we adopt {\sc atlas12} because {\sc atlas12} employs opacity sampling which allows the change of individual elemental abundances and solve for stellar atmosphere self-consistently. To better visualize the oxygen response function, we adopt the exact flux continua provided by the synthetic spectral models \footnote{The response function plotted in Fig.~\ref{fig1} adopted a normalization scheme described in \citet{ho17a} because real observed spectra do not have a priori known continua. Hence, the normalization in Fig.~\ref{fig1} is not the same as other plots in this paper.} We also convolve our model spectra to $R =2000$ to approximate the resolution of the LAMOST data.

Fig.~\ref{fig2} shows the contributions of individual species to the oxygen abundance sensitivity in the optical spectra. The middle panels show the oxygen information in typical K-giants ($T_{\rm eff} = 4800{\rm K}$, $\log g = 2.5$) which is the focus of this study. The top and bottom panels compare this information to cooler M-giants ($T_{\rm eff} = 3800{\rm K}$, $\log g = 0.5$) and hotter G dwarfs ($T_{\rm eff} = 5800{\rm K}$, $\log g = 4.5$). We assume solar metallicity throughout this paper.  The left panels show the indirect oxygen information that is not from oxygen-bearing features, and the right panels illustrate the contributions of oxygen-bearing molecules and atoms. 

For all stellar types shown in Fig.~\ref{fig2} the effect on the spectrum from atomic oxygen lines is $\ll 1$\% and is sub-dominant to all other oxygen-related effects. Moreover, for K-giants, there is negligible direct oxygen information in the optical from the oxygen-bearing molecules. The oxygen abundance is mostly imprinted on the spectra through three carbon-related molecules, namely CH, CN, and ${\rm C}_2$ \citep[see also][]{bre16}. We also tested that restricting the analysis to the wavelength range $<600\;{\rm nm}$, thereby excluding atomic oxygen features, does not change the qualitative results in Fig.~\ref{fig1}. For the cooler M-giants TiO contributes the strongest signature by an order of magnitude, as the number density of TiO molecules scale almost linearly with the O abundance. AlO and VO also contribute to the direct information at a cooler temperature but their signatures are negligible compared to TiO. There exists a wealth of both direct and indirect oxygen indicators for cool stars. For warmer stars, the information carried by molecules (both direct and indirect oxygen indicators) is greatly reduced owing to the reduced abundances of molecules in the atmospheres of hotter stars.

Fig.~\ref{fig2} shows that the optical spectrum of cool stars is sensitive to the oxygen abundance through a variety of carbon-bearing molecules.  In order to use this information to measure the oxygen abundance we must demonstrate that these features are not completely degenerate with the carbon and nitrogen abundances. That this is mathematically not the case, is shown in Fig.~\ref{fig3}. Here we have computed the C and N response functions and determined the best linear combination of these response functions that mimic the oxygen response. The red line in Fig.~\ref{fig3} shows the portion of the oxygen response function that is not covariant with C and N. This line shows that C and O can be independently measured at solar metallicity.  To understand the origin of this behavior requires an understanding of the stellar atmosphere and CNO atomic-molecular equilibrium, which is the subject of the next section.

In contrast, the oxygen information appears to be largely degenerate with carbon and nitrogen at lower metallicity (lower panel).  As a result, oxygen abundance measured with this method will have large scatter at the metal-poor regime as it is completely degenerate with C and N. Therefore, we note that while this method is useful for evaluating the bulk of the data from large surveys because they are usually more metal-rich than $[Z/{\rm H}] > -1$, the method presented in this study might not be entirely useful for the studies of globular clusters. Due to this caveat, when analyzing metal-poor stars with this method, it will be important to estimate the Cramer-Rao bound \citep[see][]{tin17a} for individual elemental abundances to provide the statistical best limit and to flag elemental abundances that have almost degenerate information.

\begin{figure*}
\centering
\includegraphics[width=1.0\textwidth]{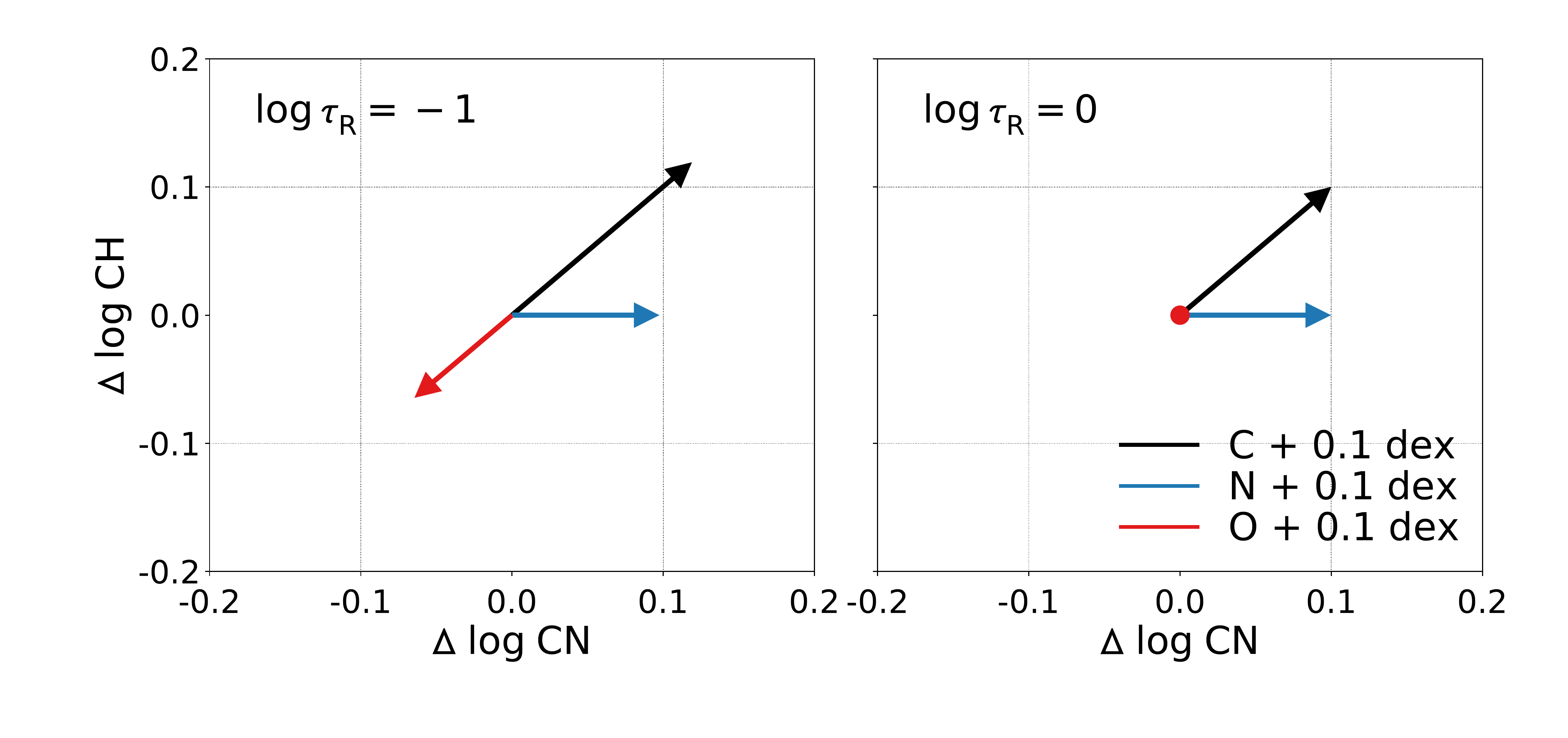}
\caption{Change in carbon-bearing molecular abundances in the stellar atmosphere, for a K-giant, in response to changes in C, N, and O abundances. Left panel shows results at a depth of $\log\,\tau_R =-1$ and right panel at $\log\,\tau_R =0$. The effect of nitrogen is restricted to the CN molecule. In the outer layers CO dominates the carbon budget and therefore strongly couples the behavior of oxygen and carbon (left panel) --- the result of which is that oxygen and carbon abundances are largely degenerate when considering the CH and CN molecules that influence the optical spectrum. At deeper layers atomic carbon dominates the carbon budget and so oxygen and carbon are no longer coupled (right panel). Figure \ref{fig4} demonstrates that the optical spectrum is sensitive to both of these regimes and so the carbon, nitrogen {\it and} oxygen abundances should be independently measurable from the carbon-bearing molecules alone.}
\label{fig5}
\end{figure*}

%
%
%
%
%
%

\section{The influence of oxygen in the stellar atmospheres of cool stars}
\label{sec:atmosphere}

The goal of this section is to understand why the carbon-bearing molecules CH, CN, and C$_2$ enable a measurement of oxygen that is not degenerate with carbon and nitrogen abundances.  To do this we must understand the structure of the stellar atmosphere and its influence on the emergent spectrum.

We begin with the two left panels in Fig.~\ref{fig4}, which show the temperature, pressure, and species abundances of key atoms and molecules as a function of the Rosseland opacity for a solar metallicity K giant ($T_{\rm eff}=4800$ K, logg$=2.5$).  There are two distinct regimes in the atmosphere: at relatively deep layers ($\log \tau_R \gtrsim 0$), the majority of C, N, and O is in atomic form.  In the outer layers ($\log \tau_R \lesssim -1$) much of the carbon is in the CO molecule and the abundances of CO and O are comparable.  As we will see below, the CO molecule plays a key role in coupling of carbon and oxygen when CO is abundant.

The right panel of Fig.~\ref{fig4} shows the approximate spectral line formation depths (for simplicity defined here as where the monochromatic line opacity, $\tau_\lambda = 1$) in units of the Rosseland opacity as a function of wavelength after smoothing to 0.3$\,$nm. The right panel reflects the combination of continuous and line opacity and shows the well-known effect that bluer wavelengths on average form in higher atmospheric layers due to the larger continuum opacity, typically larger transition probabilities, and the greater density of lines.  In this figure the line is color-coded by the ratio of CO and atomic carbon at the corresponding line formation depth.  This panel highlights two key facts - the optical spectrum is sensitive to a range of atmospheric layers, and that range encompasses both the CO-dominant and atomic carbon-dominant regimes.

In Fig.~\ref{fig5} we explore the effect of increasing the C, N, and O abundances on the number density of CH and CN molecules at two representative layers: $\log \tau_R=-1$ and 0.  At the deeper layer (right panel), where the vast majority of C, N, and O are in atomic form, the behavior of CH and CN is straightforward: increasing nitrogen by 0.1~dex results in a 0.1~dex increase in CN and no change to the other carbon-bearing molecules.  Increasing oxygen has no effect, and increasing carbon by 0.1 dex results in 0.1 dex increase in all the carbon-bearing molecules (except for C$_2$ which depends quadratically on the carbon abundance).

The effect of oxygen in the outer layer (Fig.~\ref{fig5} left panel) is qualitatively different.  Here, an increase in the oxygen abundance results in a number density {\it decrease} of all carbon-bearing molecules (except CO).  The formation of CO is energetically favourable compared to other carbon-bearing molecules due to the high dissociation energy of CO: an increase in oxygen favours the formation of more CO at the expense of molecules such as CH, CN, and C$_2$.  This behavior is not linear: a 0.1 dex increase in the oxygen abundance does not result in a 0.1 dex decrease in the CH nor CN number density.  This is because the number density of CO rivals the densities of the atomic forms of CNO.  In the higher atmospheric layers CO becomes "fuel-starved" and therefore its number density does not respond linearly to changes in only carbon or oxygen.

We can now put all of these pieces together to explain why the optical spectrum of a solar-metallicity K giant is sensitive to oxygen abundance even though there are no detectable oxygen lines.  The stellar spectrum arises from a range of atmospheric layers, including both CO-dominant (outer) and atomic-dominant (deeper) layers (Fig.~\ref{fig4}).  The redder wavelengths probing the deeper layers provide a means to measure the carbon and nitrogen abundances from the CH and CN molecular features {\it independent} of the oxygen abundance, as shown in the right panel of Fig.~\ref{fig5}.  With knowledge of the carbon abundance from the deeper layers, the behavior of CH, C$_2$, and CN at the shorter wavelengths sensitive to the outer layers then provides a measurement of the oxygen abundance (via the left panel of Fig.~\ref{fig5}).  Of course, when fitting the entire spectrum self-consistently, all of these effects are considered simultaneously and contribute to the final oxygen measurement.

The discussion above focused on solar metallicity K giants. Fig.~\ref{fig2} shows that the oxygen sensitivity is greatly diminished by 5800 K, while Fig.~\ref{fig3} shows that for metal-poor stars there is very little oxygen information that is not degenerate with carbon and nitrogen. The reason for the latter is two-fold.  First, at lower metallicity CO is less abundant, even in the outer atmospheric layers, since when CO is a minority species its number density scales quadratically with [Fe/H].  Second, owing to the lower opacity, the optical spectrum is sensitive to a narrower, and deeper range of atmospheric layers, and so the spectrum does not probe the two regimes (atomic-dominated and CO-dominated) that are important for an independent oxygen measurement.

%
%
%
%
%
%

\section{Conclusion}
\label{sec:conclusion}

We set out to to provide a physical explanation for the result from \citet{tin17b} that oxygen abundances can be measured from low resolution ($R \simeq 2000$) optical spectra of metal-rich red giants, even though there are no detectable oxygen features in the spectra.  We have shown that through CNO atomic-molecular equilibrium, and in particular the CO molecule, the abundance of oxygen can affect the strength of CH and CN molecular features, which contain numerous strong transitions throughout the optical spectrum.  This signal is not degenerate with C or N abundance variations for ${\rm [Z/H]}\,>-1$, because the optical spectrum is sensitive to a range of atmospheric layers including both CO-dominant and atomic-dominant regions.  

This approach of measuring an elemental abundance, in this case oxygen, from spectral features that do not involve oxygen may seem considerably more ``indirect'' than ``directly" measuring one or more individual atomic absorption lines of the element. But within the framework of fitting spectra with physical models it is not really more indirect: any modeling must rely on atomic data and an atmosphere model that correctly captures the thermodynamic state and composition as a function of (optical) depth, along with the equation of radiative transfer. Without these modeling ingredients, individual atomic lines are also not quantitatively interpretable; with them, the predictions for the spectrum as outlined here are just as direct. Perhaps the most important difference of the approach described here is that it must rely on the simultaneous fitting of all spectral labels, which however is possible \citep{tin17b}.

In good part the approach we describe here works well for oxygen because it is such an abundant element that abundance changes manifestly affect other species. But the more broad approach is not restricted to oxygen: for example, the abundance of an effective electron-donating element (e.g., Mg, Ca) may affect -- by way of the photosphere's ionization structure -- many parts of the spectrum beyond the wavelengths of atomic lines from the element itself. Therefore, this somewhat unorthodox approach to measuring individual element abundances should have applications beyond the case laid out here.

%
%
%
%
%
%

\section*{Acknowledgments}
\label{sec:acknoledgements}
We thank Robert Kurucz for providing his stellar atmosphere and spectrum synthesis codes and atomic and molecular line lists and Fiorella Castelli for allowing us to use her linux versions of the programs, without which this work would not be possible. YST is supported by the Carnegie-Princeton Fellowship, the Martin A. and Helen Chooljian Membership from the Institute for Advanced Study in Princeton and the Australian Research Council Discovery Program DP160103747. CC acknowledges support from NASA grant NNX13AI46G, NSF grant AST-1313280, and the Packard Foundation. HWR's research contribution is supported by the European Research Council under the European Union's Seventh Framework Programme (FP 7) ERC Grant Agreement n. [321035] and by the DFG's SFB-881 (A3) Program. MA acknowledges funding from the Australian Research Council (FL110100012, DP150100250, CE170100013).

%
%
%
%
%
%

\end{CJK*}

\bibliography{biblio.bib}

\end{document}